\documentclass[preprint]{elsarticle}
\usepackage{graphicx}
\usepackage{epstopdf}
\usepackage{color}

\journal{Physica A}

\begin{document}

\begin{frontmatter}

\title{Social consensus and tipping points with opinion inertia}

\author{C. Doyle\corref{null} $^{1,2}$}

\author{S. Sreenivasan\corref{null} $^{1,2,3}$}

\author{B. K. Szymanski\corref{null} $^{2,3}$}

\author{G. Korniss\corref{c1} $^{1,2}$}
\ead{korniss@rpi.edu}

\address{$^1$ Dept. of Physics, Applied Physics and Astronomy, Rensselaer Polytechnic Institute
110 Eighth Street, Troy, NY 12180}
\address{$^2$ Network Science and Technology Center, Rensselaer Polytechnic Institute
110 Eighth Street, Troy, NY 12180}
\address{$^3$ Dept. of Computer Science, Rensselaer Polytechnic Institute
110 Eighth Street, Troy, NY 12180}
\cortext[c1]{Corresponding author}

\begin{abstract}
When opinions, behaviors or ideas diffuse within a population, some
are invariably more {\it sticky} than others. The stickier the
opinion, behavior or idea, the greater is an individual's inertia to
replace it with an alternative. Here we study the effect of
stickiness of opinions in a two-opinion model, where individuals
change their opinion only after a certain number of consecutive
encounters with the alternative opinion. Assuming that one opinion
has a fixed stickiness, we investigate how the critical size of the
competing opinion required to tip over the entire population varies
as a function of the competing opinion's stickiness. We analyze this
scenario for the case of a complete-graph topology through
simulations, and through a semi-analytical approach which yields an
upper bound for the critical minority size. We present analogous
simulation results for the case of the Erd\H{o}s-R\'enyi random
network. Finally, we investigate the coarsening properties of sticky
opinion spreading on two-dimensional lattices, and show that the
presence of stickiness gives rise to an effective surface tension
that causes the coarsening behavior to become curvature-driven.
%We focus on the scenario
%where initially, a minority of the population adopts an opinion that
%is as sticky or stickier than that of the majority, and investigate
%how the critical size of the initial minority required to tip the
%entire population over to its opinion, depends on the stickiness of
%the minority opinion.
\end{abstract}


\end{frontmatter}

%\maketitle

\section{Introduction}
%%%%%%%%%%% SS INTRO
Social networks represent potent structures on which opinions and
behavior diffuse, and on which tipping points \cite{Gladwell2002} in
the adoption of opinions and behavior arise. A number of theoretical
studies investigating the diffusion of ideas, opinions, or behavior,
have focussed on understanding how a small fraction of initiators
\cite{Watts2002,Watts2007,Singh2013} or committed proselytizers
\cite{Galam2007,Xie2011,Zhang_Chaos2011,Xie_PLOSONE2011,Turalska_SREP2013,Waagen_PRE2015}
of an idea can tip over the entire network to adopt the same.
Furthermore, within these studies, various sources of competition to
the spread of an idea have been considered - for example, a
competing idea that is spreading over the network, or a bias
external to the network that is trying to suppress the spread of an
idea. More pertinently, however, the tendency of individuals
themselves to be pliable to change is dynamic and could be dependent
on their activity. In particular, individual behavior itself is
subject to some inertia that opposes any change in the beliefs or
opinions adopted by the individual \cite{Nickerson98}. A well known
example of such inertia is the phenomenon of {\it confirmation bias}
in social psychology, where individuals tend to favor beliefs that
conform to their currently held position. Overcoming such individual
inertia to change is therefore a primary consideration in campaigns
for public opinion change \cite{Heath2007}.

Motivated by this phenomenon, we study a theoretical model of
opinion change where individual opinion change depends on the
current state of the individual as well as the recent history of the
opinions she has encountered in interactions with her neighborhood.
Specifically, we assume that there are two opinions vying for
adoption on a social network, and each individual requires a
pre-defined threshold number of interactions with the alternative
opinion, before switching to it. Thus each opinion is {\it sticky} to its respective extent \cite{Heath2007}. Furthermore, in an attempt to
capture the effect of confirmation bias, we posit that an
individual's memory of a stream of encounters with the alternative
opinion is erased by a single interaction in which he encounters his
currently held opinion.  There is some precedent to studying such a
memory-based model of switching between states. Dodds and Watts
\cite{Dodds2005} studied a model of disease contagion where a
susceptible person became infected only when his interactions with
infected neighbors within a certain prior time window had led to a
pre-defined infection-dosage threshold being exceeded.  More
pertinently to the current study, Dall'Asta and Castellano
\cite{DallAsta2007} studied a variant of the Naming Game with two
pure opinions, where an individual switches to the intermediate
state only when the number of times he has encountered the opposing
opinion exceeds some pre-defined threshold. Our model thus is a
special case of  \cite{DallAsta2007} where the memory window is
exactly equal to the threshold, and where no intermediate state is
present. In contrast to the work done in \cite{DallAsta2007}, here
our focus is to look at the fraction of initiators required to bring
about a tipping point. The effect of stickiness has also been
studied in the context of the Naming Game in \cite{Baronchelli2007}
and more recently in \cite{Thompson2014}. In these studies, the
stickiness parameter quantifies the probabilities with which a node
in a mixed-opinion state rejects a pure state that it
encounters in an interaction with its neighbors. The introduction of
the stickiness parameter for nodes in the mixed-opinion state, gives
rise to a phase transition between a regime where the consensus
states are stable (when stickiness is low) to one where the
consensus states are unstable and the system gravitates to a stable
state with a non-zero density of mixed-opinion nodes.

A recent work \cite{Cui2014} has studied a variant of the SIR model
where the infection probability is a function of the number of
infectious neighbors, and parametrized by two parameters that they
designate as stickiness and persistence. Despite the nomenclature,
the term stickiness is utilized in \cite{Cui2014} to designate the
slope of the infection probability of a susceptible node as a
function of the size of its infected neighborhood, and therefore
bears little similarity to the context that we study. Finally, recent empirical
findings \cite{Galehouse2014} demonstrate the dependence of social network properties on cultural attributes of the population, suggesting
that stickiness could also be similarly influenced by cultural factors.

\section{Description of model}
Here we define the microscopic rules of our model. We assume that
every individual on a social network initially adopts one of two
opinions, which we designate $A$ and $B$. The fundamental mechanism
in our model for the change in individual states is the interaction
of pairs of individuals, which represent speaker-listener pairs. In
each such interaction, the speaker conveys his opinion to the
listener, and in response to this conveyed opinion, the listener
changes his state or continues to hold the same state depending on
the rules of the model. We elaborate on these rules in the next few
lines. First, each opinion has a pre-defined {\it stickiness},
designated as $w_A$ and $w_B$ respectively. The stickiness of an
opinion represents the inertia present in an individual adopting
that opinion, to change her state. In terms of the model, the
stickiness $w_A$ ($w_B$) of an individual in state $A$ ($B$) is the
number of consecutive times she requires to hear the opinion $B$
($A$), before she switches her opinion to $B$ ($A$). Thus, we can
assume that each individual keeps a counter dedicated to counting
the number of times she encounters the alternative opinion, which
resets to zero either when the required number of consecutive
interactions of the alternative opinion are heard, or whenever the
current opinion is heard. Note that in the former case, the counter
also switches the opinion that it is keeping track of.
%%%%%%%%%%%%%%%%%%%%%%%%%%%%%%%%%%%%%%%%%%%%%%%%%%%%%%%%%%%%%%%%%%%%%%%%%
In our current model implementation, we assumed that
exposure to a different/same opinion only impacts the individuals'
counter when their role is the listener in a pairwise interaction.
Naturally, one may consider the scenario where both the speaker's
and listener's counters are affected by the interactions (i.e., the
speaker can also be reinforced in her view). We did some
explorations on this generalization of the model, and have found
that there are no qualitative differences in the results.
%%%%%%%%%%%%%%%%%%%%%%%%%%%%%%%%%%%%%%%%%%%%%%%%%%%%%%%%%%%%%%%%%%%%%%%%%%%%

In summary, the model dynamics proceeds as follows. The individuals
(nodes) in the network are initially assigned one of the two
opinions such that we have prescribed fractions $p_A$ and $p_B = 1-p_A$ of
nodes in states $A$ and $B$ respectively. Then at each microscopic
time step,  a random node is chosen from the system and designated
as the speaker.  A random node is selected from among the speaker's
neighbors and designated as the listener.  If the listener's opinion
is the same as the speaker's, it's progress towards switching is
reset to zero.  If the listener's opinion is different from the
speaker's, the listener's count towards switching increases by one.
If the listener's count becomes equal to it's opinion's stickiness,
it adopts the alternative opinion and begins a fresh count. We
assume that $N$ such microscopic time steps constitute unit time,
where $N$ is the network size. Thus, the event that a node is
selected as a speaker is a Poisson process with rate $1$.

\section{Results and Discussion}
\subsection{Complete Graph}

\begin{figure}
%\vspace{-.75cm}
\center
\includegraphics[scale=0.75]{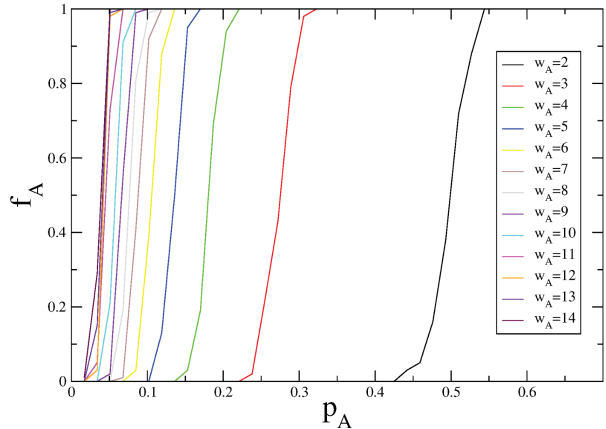}
\vspace{-0.50cm}
\caption{The fraction of simulation runs on a complete graph with $N$$=$$1000$ that reach consensus on opinion $A$ vs. the initial population fraction of opinion $A$ for different stickiness values
of the $A$ opinion. For these simulations $w_B=2$.}
\label{figscurve}
\end{figure}

\begin{figure}
\center
\includegraphics[scale=.61]{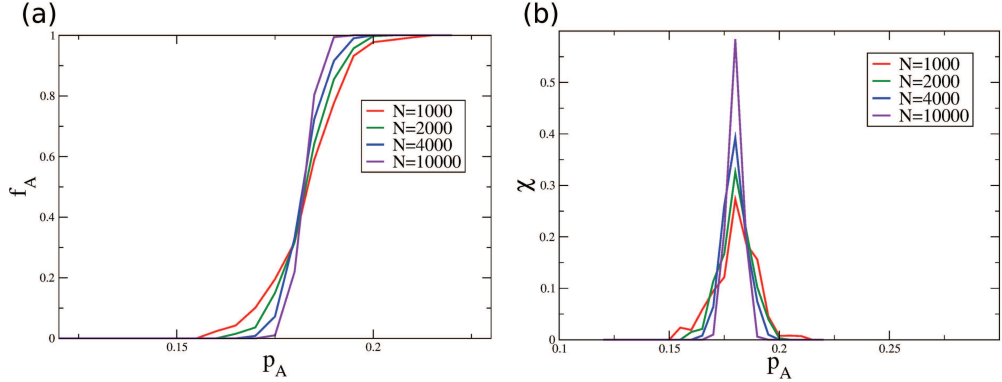}
\caption{(a) The fraction of runs reaching consensus on opinion $A$ vs. the
initial population fraction of opinion $A$ for different system sizes.
These runs have $w_A=4, w_B=2$, and are averaged over 1000 runs.
(b) Forward derivative of the fraction of runs reaching consensus on opinion $A$ from (a) vs.
the initial population fraction of opinion $A$.}
\label{figsderiv}
\end{figure}

First, we investigate the outcome of these rules on a complete graph
through Monte-Carlo simulations. As shown in Fig.~\ref{figscurve},
we vary the fraction, $p_A$, of nodes adopting opinion $A$, and
measure the fraction of simulation runs (over a total of $500$ runs)
for which the system reaches consensus on opinion $A$. We keep the
stickiness of the opinion $B$ fixed at $w_B = 2$ and vary the
stickiness $w_A$ of opinion $A$. For a finite system, for every value of $w_A$, the
fraction of runs reaching consensus on opinion $A$, $f_{A}$, follows
a typical S-shaped curve [Fig.~\ref{figscurve}].
%%%%%%%%%%%%%%%%%%%%%%%%%%%%%%%%%%%%%%%%%%%%%%%%%%%%%%%%%%%%%%%%%%%%%%
For increasing system sizes, these curves are becoming
progressively sharper [Fig.~\ref{figsderiv}(a)], approaching a
discontinuous transition in the infinite system-size limit and
indicating the existence of a tipping point at a critical fraction
$p_c$.
%%%%%%%%%%%%%%%%%%%%%%%%%%%%%%%%%%%%%%%%%%%%%%%%%%%%%%%%%%%%%%%%%%%%%%%
%We identify $p_c$ as the value of $p_A$ at which the fraction of
%runs reaching consensus on $A$ becomes equal to $1/2$.
For a finite system size $N$, we identify $p_c$ where the (forward)
derivative of the fraction of runs reaching $A$-consensus,
$\chi\equiv df_{A}/dp_{A}$, is maximum [Fig.~\ref{figsderiv}(b)].
These results also indicate that the finite-size effects of the
location of the critical point are negligible for this transition.
%%%%%%%%%%%%%%%%%%%%%%%%%%%%%%%%%%%%%%%%%%%%%%%%%%%%%%%%%%%%%%%%%%%%%

As demonstrated by the results shown in Fig.~\ref{figscurve}, for values of $w_A
\geq 3$, at the critical point, the opinion $A$ initially
constitutes the minority opinion. Thus, having a stickiness even
marginally greater than that of the majority opinion allows the
minority opinion to tip over the entire population, as long as the
minority fraction is greater than $p_c$. For equal stickiness $w_A =
w_B = 2$, the fraction of opinion $A$ holders must from the start be
the majority opinion, in order to win over the population.

\begin{figure}
\center
\includegraphics[scale=.61]{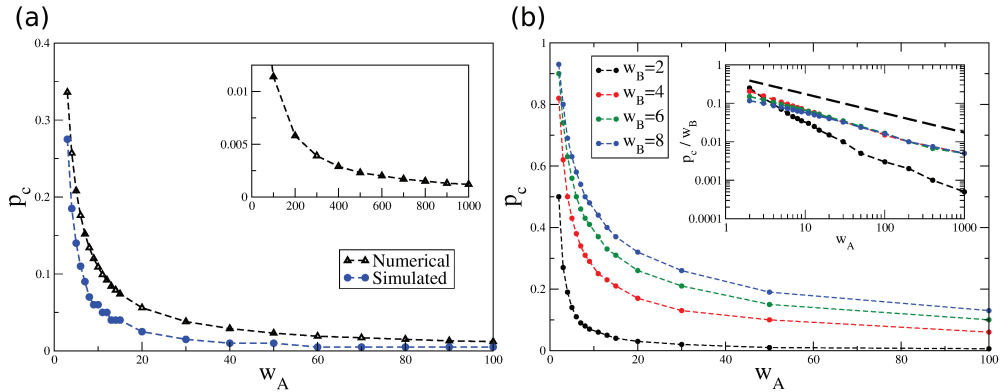}
\caption{(a) Comparison between the values of the critical fraction
$p_c$ as a function of the stickiness of opinion A for a complete-graph with $N$$=$$1000$ obtained from
simulations (averaged over 100 runs) and through the semi-analytic approach (Sec.~3.2). In both cases $w_B=2$.
The inset shows the extended numerical results [using Eqs.~(\ref{evolnAnB}) and (\ref{evolsa0sb0}) up to values of $w_A=1000$.
(b) Critical populations obtained by simulations for different values of $w_B$ with $N=1000$ and averaged over 100 runs.
The inset shows the scaled data on a log-log scale. The dashed line, for reference, corresponds to a power law with
exponent $-1/2$.}
\label{figanalytical}
\end{figure}

As the stickiness of the minority opinion is increased, the takeover
of the entire network occurs at progressively smaller minority
fractions. As shown in Fig.~\ref{figanalytical}, $p_c$ appears to
converge to zero as $w_A \to \infty$. For simulations shown here, $N
= 1000$, and hence the smallest value that $p_c$ can adopt is
$0.001$. However, we show using a semi-analytic approach (Sec.~3.2)
that an upper bound to the critical value $p_c$ itself converges to
$0$ as $w_A \to \infty$ [Fig.~\ref{figanalytical}(a) inset], which
confirms that the critical fraction vanishes for asymptotically
large stickiness.
%%%%%%%%%%%%%%%%%%%%%%%%%%%%%%%%%%%%%%%%%%%%%%%%%%%%%%%%%%%%%%%%%%%%%%%%%%%%

%%%%%%%%%%%%%%%%%%%%%%%%%%%%%%%%%%%%%%%%%%%%%%%%%%%%%%%%%%%%%%%%%%%%%%%%%%%%%%
Note that the full dependence of the tipping point
$p_{c}$ on the stickiness parameters $w_{A}$ and $w_{B}$ is rather
complex and non-linear. Our simulation results and scaling suggest that
%%%%%%%%%%%%%%%%%%%%%%%%%%%%%%%%%%%%%%%%%%%%%%%%%%%%%%%%%%%%%%%%%%%%%%%%%
\begin{equation}
p_c\simeq \frac{\rm const.}{(w_{A}^{1/2}/w_{B})} \propto w_{B} w_{A}^{-1/2} \;,
\end{equation}
%%%%%%%%%%%%%%%%%%%%%%%%%%%%%%%%%%%%%%%%%%%%%%%%%%%%%%%%%%%%%%%%%%%%%%%%%
in the $1\ll w_{B}\ll w_{A}$ limit, as shown in the inset of Fig.~\ref{figanalytical}(b).
%%%%%%%%%%%%%%%%%%%%%%%%%%%%%%%%%%%%%%%%%%%%%%%%%%%%%%%%%%%%%%%%%%%%%%%%%%%%%%%%

\subsection{Analytical approximation for critical fraction on complete graphs}
For convenience, in this subsection we denote the stickiness of opinion $A$ by $w$ and the stickiness of opinion $B$ by $v$. We denote the fraction of nodes holding opinions $A$ and $B$ by $n_A$ and $n_B$ respectively. The fraction of nodes holding opinion $A$ is comprised of distinct subpopulations that hold opinion $A$ and have accrued a certain number of consecutive {\it hits} from opinion $B$. We denote the fractional sizes of these subpopulations by $s_{a,0},s_{a,1}, \cdots, s_{a,w-1}$ respectively. Analogously, the subpopulations for opinion $B$ are denoted as $s_{b,0},s_{b,1},\cdots, s_{b,v-1}$ respectively. Thus
\begin{eqnarray}
n_A & = & \sum_{x=0}^{x=w-1} s_{a,x}  \nonumber \\
n_B & = & \sum_{x=0}^{x=v-1} s_{b,x}.
\label{totalpop1}
\end{eqnarray}

We can write evolution equations for the density of nodes in states $A$ and $B$, by noting that a change in opinion occurs when a node whose counter for the alternate opinion is just below the stickiness of its current opinion, encounters the alternate opinion. Thus, the density of nodes with opinion $A$ evolves according to the equation:
\begin{equation}
\frac{dn_A}{dt} = -n_B ~  s_{a,w-1} + n_A ~ s_{b,v-1},
\label{evolnA}
\end{equation}
where the first term captures the loss of nodes in state $A$, resulting from nodes represented by the fraction $s_{a,w-1}$ hearing opinion $B$. The second term analogously captures the gain resulting from nodes represented by the fraction $s_{b,v-1}$ hearing opinion $A$. Similarly,
\begin{equation}
\frac{dn_B}{dt} = -n_A ~ s_{b,v-1} + n_B ~ s_{a,w-1},
\label{evolnB}
\end{equation}
Next, in order to make these equations tractable, we introduce a quasi-steady state approximation for obtaining the subpopulation fractions for each opinion. Specifically, we assume:

\begin{eqnarray}
s_{a,x} = s_{a,0}  ~ (n_B)^{x}    \nonumber \\
s_{b,x} = s_{b,0}  ~ (n_A)^{x}.
\end{eqnarray}

Namely, we assume that the fraction of nodes in state $\{a,x\}$ at a given time is approximately equal to the probability, given the systems current state, of picking a node in state $\{a,0\}$, and picking a node in state $B$ on every one of $x$ trials with replacement. This assumption, commonly used in the study of chemical reaction systems with intermediates, is known as the quasi-steady-state assumption \cite{Briggs1925}, referring to the fact that the intermediate subpopulations arising in the transition from state $\{a,0\}$ to state $\{b,0\}$ and vice-versa, are assumed to be in steady-state. This can be seen from the evolution equation for a particular subpopulation, say $\{a,x\}$:
\[
\frac{ds_{a,x}}{dt}  = -n_A  ~ s_{a,x} - n_B  ~ s_{a,x} + n_B  ~ s_{a,x-1}
\]
Since $n_A + n_B = 1$, the steady-state expression for fraction of nodes in state $\{a,x\}$ is: $s_{a,x} = n_B  ~ s_{a,x-1}$. Thus, $s_{a,x} = (n_B)^x ~  s_{a,0}$.
Using this approximation, Eqs.~(\ref{evolnA}), (\ref{evolnB}) become:
\begin{eqnarray}
\frac{dn_A}{dt} &=& -(n_B)^w ~  s_{a,0} + (n_A)^v  ~ s_{b,0} \nonumber \\
 \frac{dn_B}{dt} &=& -(n_A)^v  ~ s_{b,0} + (n_B)^w ~ s_{a,0}
 \label{evolnAnB}
\end{eqnarray}
Additionally, we have evolution equations for nodes in states $\{a,0\}$ and $\{b,0\}$ as well:
\begin{eqnarray}
\frac{ds_{a,0}}{dt}=-n_B  ~ s_{a,0} + n_A ~ \sum_{x=1}^{w-1} s_{a,x} + n_A ~ s_{b,v-1} \nonumber \\
\frac{ds_{b,0}}{dt}=-n_A ~ s_{b,0} + n_B ~ \sum_{x=1}^{v-1}s_{b,x} + n_B ~ s_{a,w-1}
\end{eqnarray}
and using the quasi-steady-state assumption and Eqs.~(\ref{totalpop1}), we obtain:
\begin{eqnarray}
\frac{ds_{a,0}}{dt}=-n_B ~ s_{a,0} + n_A ~ (n_A-s_{a,0}) + (n_A)^v  ~ s_{b,0} \nonumber \\
\frac{ds_{b,0}}{dt}=-n_A ~ s_{b,0} + n_B ~ (n_B - s_{b,0}) + (n_B)^w  ~ s_{a,0}
\label{evolsa0sb0}
\end{eqnarray}
We numerically solve the coupled equations, Eqs.~(\ref{evolnAnB})
and ~(\ref{evolsa0sb0}) for different initial fractions $n_A^{\rm
init},n_B^{\rm init}$ (with $n_A^{\rm init} < n_B^{\rm init}$) and
obtain the steady state values of $n_A$ and $n_B$ respectively. We
then record the smallest value of  $n_A^{\rm init}$ at which the
steady state value of $n_A$ becomes greater than $0.99$ and
designate this as the critical initial minority fraction $p_c$
required to tip the system over. Figure~\ref{figanalytical} shows a
comparison for the tipping point $p_c$ obtained through this
semi-analytical approach and that obtained from simulation for
different stickiness values of opinion $A$ (while the stickiness of
the other opinion is held fixed at $w_B$$=$$2$). The cause of the
higher $p_c$ values yielded by the semi-analytical approach is the
overestimation of subspecies densities ($s_{a,x},s_{b,x}$ for $x
> 0$) in the initial phase of the dynamics - in reality the
subspecies densities take some length of time to attain non-zero
values. This overestimation favors the sustenance of nodes in state
$B$, since they are initially in the majority. As a result, the
fraction of nodes in state $A$ required to tip the system over, as
estimated by the quasi-steady-state approximation, is larger. Thus,
the semi-analytical estimate of $p_c$ consistently represents an
upper-bound to the value observed in simulations.  Furthermore, in
the event that the stickiness of opinion $A$ diverges,
Eq.~(\ref{evolnAnB}) shows that for any non-zero initial density of
$A$ opinions, $n_A$ grows monotonically while $n_B$ decays
monotonically, showing that the true critical fraction $p_c$ is
bounded above by a value that vanishes in the asymptotic limit of
stickiness.

%There is a small deviation between the analytically obtained values and those obtained from simulations which is likely a manifestation of the quasi-steady-state approximation. However, the analytical value consistently represents an upper-bound to the value observed in simulations. Furthermore, in the event that the stickiness of opinion $A$ diverges, Eq.~(\ref{evolnAnB}) shows that for any non-zero initial density of $A$ opinions, $n_A$ grows monotonically while $n_B$ decays monotonically, showing that the true critical fraction $p_c$ is bounded above by a value that vanishes in the asymptotic limit of stickiness. XXX ADD STATEMENT HERE.

\subsection{Erd\H{o}s-R\'enyi random graph}
\begin{figure}[!t]
\center
\includegraphics[scale=.44]{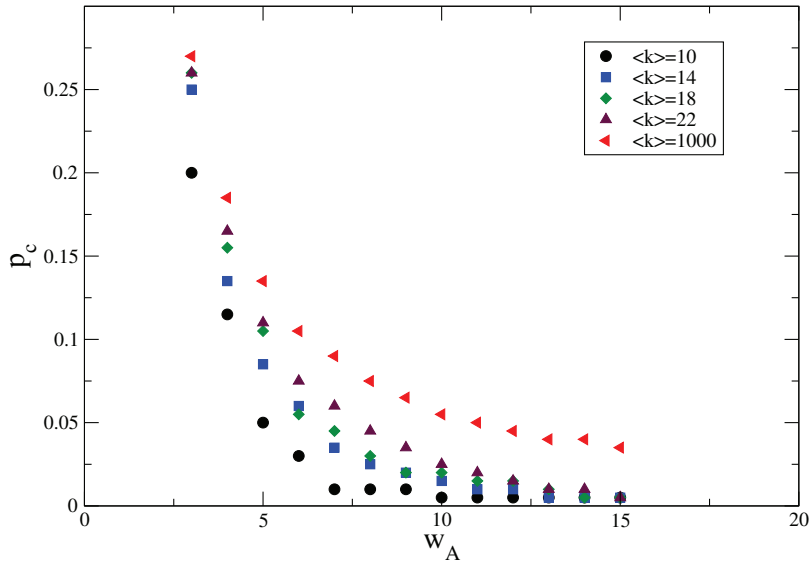}
\caption{The critical fraction $p_c$ vs. the stickiness of opinion A on Erd\H{o}s-R\'enyi
graphs with $N$$=$$1000$ for various average degree $\langle k \rangle$.}
\label{figpoisson}
\end{figure}

A similar asymptotic dependence of $p_c$ on $w_A$ with $w_B=2$ is observed for Erd\H{o}s-R\'enyi random graphs of size $N=1000$, as shown in \ref{figpoisson}.  Lowering the average degree of the graph $\langle k \rangle$ tends to lower the critical value. For comparison, we also show the critical values obtained for the corresponding complete graph with $1000$ nodes.

\subsection{Curvature-driven coarsening in the presence of stickiness}
\begin{figure}[!t]
\center
\includegraphics[scale=.50]{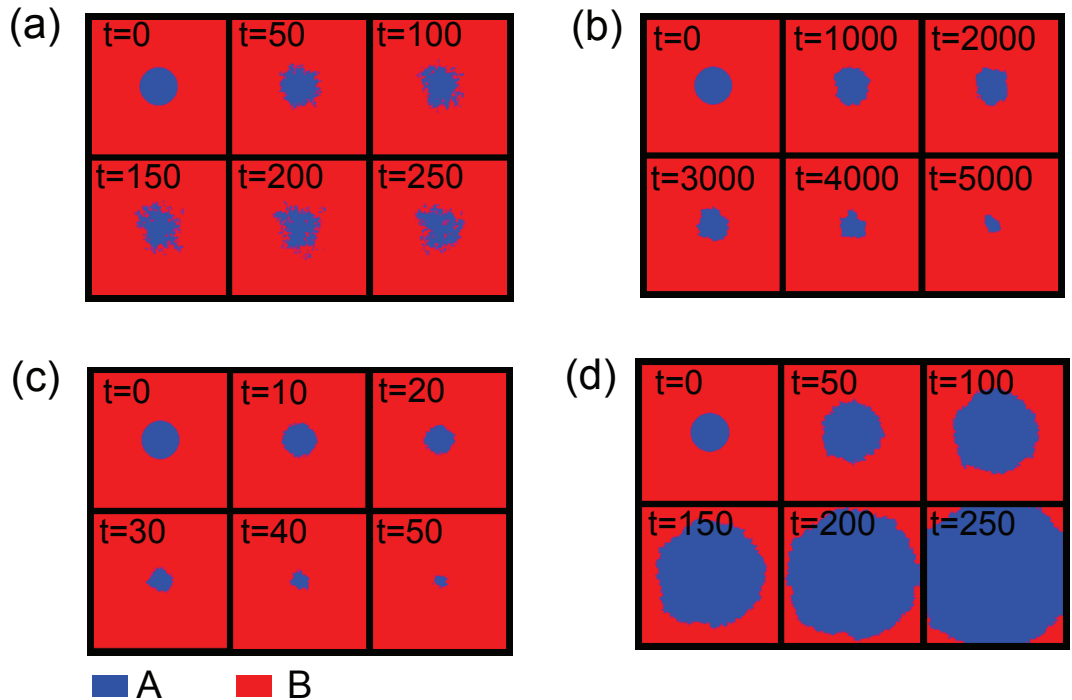}
\caption{ Snapshots of the evolution of a droplet of opinion $A$ nodes in a sea of $B$ nodes under different combinations of stickiness parameters. The nodes occupy the sites of a $250 \times 250$ 2D square lattice without periodic boundaries. Opinion $A$ is in the minority in every case and represented  in blue. Nodes with opinion $B$ are colored red.  (a) Without stickiness i.e. $w_A = w_B = 1$, the model becomes identical to the voter model, and consistent with observations for the latter, the interface roughens diffusively, without any perceivable surface tension. With the introduction of stickiness in at least one of the two opinions, (b) $w_A = w_B=2$, (c) $w_A = 1, w_B = 2$, (d) $w_A = 2, w_B = 1$, the interface evolution becomes curvature driven, and the droplet retains its roughly circular shape as it grows or decays.}
\label{figdroplet}
\end{figure}

Next, we investigate how the introduction of stickiness into the rules of opinion change affects the coarsening behavior of the system. To facilitate comparison with previous studies, we investigate the evolution of a circular droplet of nodes in state $A$ immersed in a sea of nodes holding opinion $B$ in two dimensions.  First, we visually inspect this evolution under the rules of the model, for various combinations of values for $w_A$ and $w_B$. The nodes in this case, are the sites of a square lattice (with each node connected to $4$ nearest neighbors) of side $L=250$ without periodic boundary conditions. The droplet initially has a radius of $R_0=35$. For $w_A = w_B = 1$, there is no stickiness in either opinion, and the dynamics reduces to that of the voter model \cite{Liggett1999}, where one interaction with the alternative opinion is sufficient to cause a node to change its opinion. Figure~\ref{figdroplet}(a) shows the evolution of the droplet in this case. As demonstrated in a previous study \cite{Dornic2001}, the noise-driven roughening of the interface is clearly visible as the droplet evolves.
Next, we introduce stickiness in the opinions by assuming $w_A = w_B= 2$. The initial conditions are identical to those for the case shown in Fig.~\ref{figdroplet}(a). Figure~\ref{figdroplet}(b) shows a markedly different picture and the presence of an effective surface tension in the model is evident from the preservation of interface smoothness over time. This curvature-driven evolution is consistent with behavior observed in prior studies on voter-like models with intermediate states \cite{DallAsta2008,Vazquez2008,Zhang2014} or memory \cite{DallAsta2007}, since the effect of stickiness (or memory) is similar to that of intermediate states that intercede the transition between two opinions.
Finally, stickiness in only one of the two opinions is sufficient (see Figs.~\ref{figdroplet}(c),(d)) to keep curvature-driven behavior intact.

\begin{figure}[!t]
\center
\includegraphics[scale=.60]{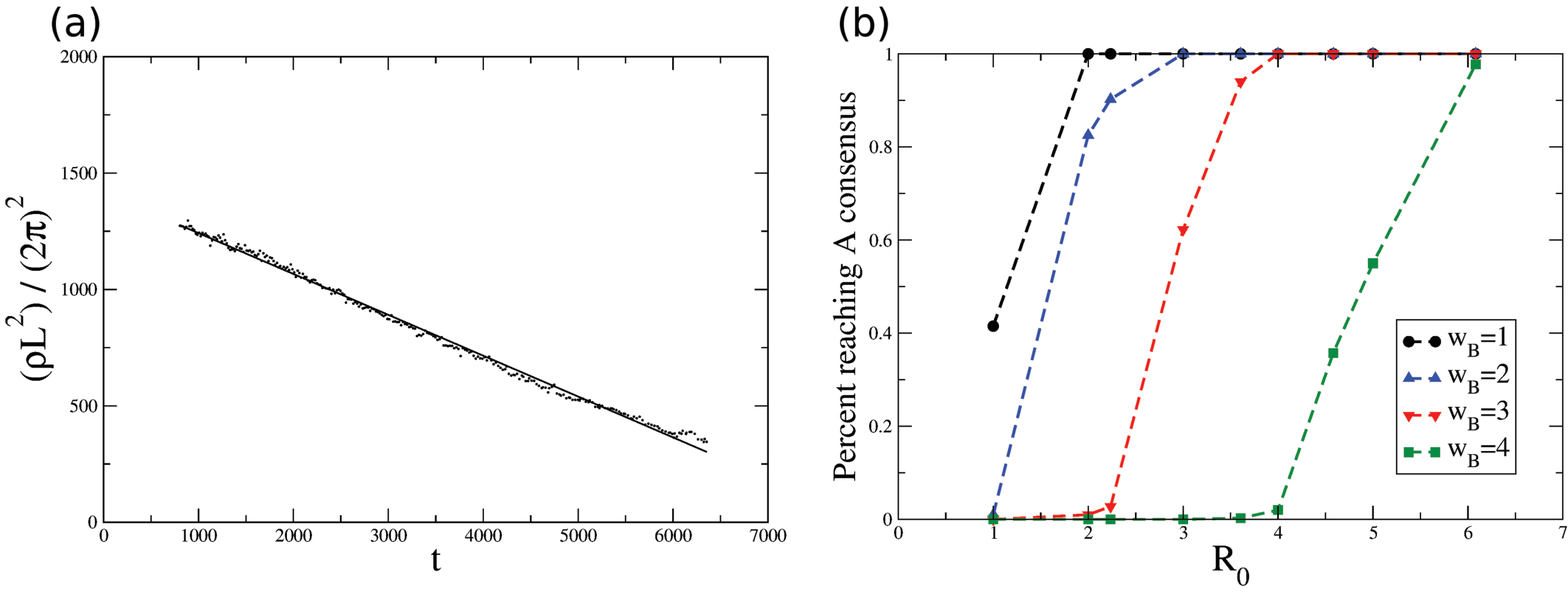}
\caption{(a) The radius of a circular droplet of opinion $A$ nodes in a sea of $B$ of nodes as a function of time for $w_{A}=w_{B}=2$.
The radius is expressed in terms of the interface density $\rho$ and the lattice size (linear dimension of the 2D square lattice) $L=250$,
and shows a linear decrease with time.
(b) The growth or decay of the circular droplet depends on its initial radius. Shown here is the fraction of simulation runs where the initial droplet grows and takes over all sites on the lattice. In these simulations, the stickiness for opinion $A$ is held fixed at $w_A = 5$ and the square lattice has dimensions $35 \times 35$. }
\label{figcurvaturedriven}
\end{figure}

Next, we investigate the coarsening behavior quantitatively. We track the evolution of the density of interfaces, $\rho(t)$ i.e. the fraction of nearest-neighbor pairs which differ in their opinion. This is a commonly used order parameter that characterizes the coarsening behavior \cite{Castellano2003, Castellano2005}.  For curvature-driven coarsening systems, the radius of the droplet changes linearly with time \cite{Bray1994}. In 2D, it follows that the interface density also grows or decays linearly in time i.e. $\rho(t) \sim  c_1 \pm c_2~t$, where $c_1$ and $c_2$ are constants. Whether the droplet grows or decays depends on both the initial size of the droplet, as well as the values of stickiness for the two opinions.
As shown in Fig.~\ref{figcurvaturedriven}(a), the decay in interface density is indeed linear, as predicted by theory. Here, the initial radius of the droplet is $R=35$, the lattice size is $L=250$, and the stickiness parameters are $w_A=w_B=2$. Figure~\ref{figcurvaturedriven}(b) shows the fraction of simulation runs (over a total of $400$ runs) for which the droplet grows and spreads over the entire lattice (with $L=35$) as a function of the initial droplet radius for various combinations of stickiness. The results indicate the existence of a critical initial droplet radius for every combination, such that the probability of droplet growth sharply rises for initial radii above this critical value \cite{Bray1994}.

\begin{figure}[!t]
\center
\includegraphics[scale=1.10]{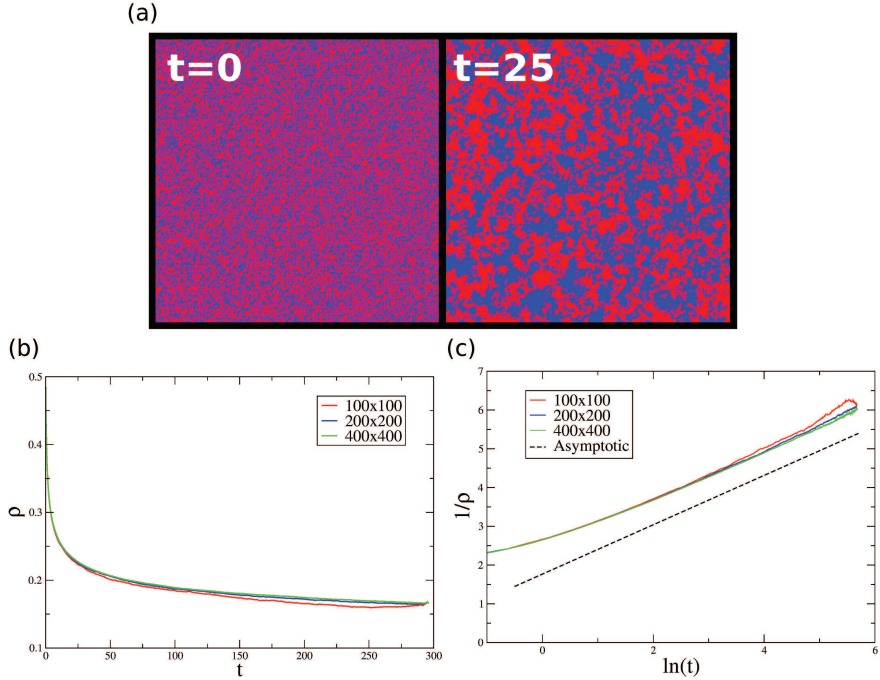}
\caption{
(a) Snapshots of the evolution of a system from random initial condition for $w_A=w_B=1$ (becoming equivalent to the voter model).
The color code is the same as in Fig.~\ref{figdroplet}. The lattice is a $100 \times 100$
2D square lattice with open boundary conditions.
(b) The interface density $\rho$ as a function of time $t$
on a 2D lattice with $w_A=w_B=1$ for various system sizes.
(c)  The same simulation data as in (b) but plotted as the inverse interface
density vs logarithmic time in order to compare to the exact
asymptotic limit of the voter model,
$1/\rho\simeq(2/\pi)\ln(t)+\ln(256)/\pi$ (dashed line) \cite{Frachebourg1996}.}
\label{fignoisedriven}
\end{figure}

In diffusive systems like the voter model, it has been theoretically
demonstrated that in the asymptotic long-time limit, the interface density decays logarithmically
\cite{Frachebourg1996} in 2D, under random initial
conditions (see details below). Figure~\ref{fignoisedriven}(a) shows two snapshots of
the coarsening process for the case $w_A = w_B = 1$, on a $100
\times 100$ 2D square lattice at time $t=0$ (random initial
conditions) and at $t=25$, respectively. The diffusive nature of
interface evolution, characteristic of the voter model, is clearly
visible and is consistent with the behavior observed in the evolution of
the circular droplet shown in Figure~\ref{figdroplet}(a).

Figure~\ref{fignoisedriven}(b) shows the slow decay of
the interface density as a function of time. One must be careful,
however, as the exact asymptotic inverse {\em logarithmic}
dependence of the interface density on time has long been known to
be challenging to demonstrate numerically
\cite{Frachebourg1996,Evans1993}. Specifically, for the voter model,
the leading-order asymptotic behavior for the interface density is
$\rho\simeq\pi/[2\ln(t)+\ln(256)]$ \cite{Frachebourg1996}. As
indicated by the results of our simulations
[Fig.~\ref{fignoisedriven}(c)], our model with $w_A = w_B = 1$
approaches (albeit slowly) precisely this type of long-time
asymptotic behavior, as expected.

\section{Conclusions}
We have modeled a scenario where two competing opinions, ideas or
behaviors vie for adoption in a social network. Each opinion is
endowed with an inherent stickiness that impedes an individual
adopting that opinion from switching to the alternative opinion.

We have demonstrated that the stickiness of the dominant opinion on
a social network determines how large the fraction of minority
opinion holders needs to be in order to tip over the population to
the initially minority opinion. We have further shown that
increasing the stickiness of the minority opinion lowers the
critical fraction required for its mass adoption dramatically as
shown in Fig.~\ref{figanalytical}. In practical contexts, the
stickiness of an opinion or behavior is related to the costs
incurred, or incentives provided by its adoption, in comparison with
the alternative. On two-dimensional lattices, we have shown that the
presence of stickiness in just one of the two opinions causes the
system's behavior to belong to the universality class of models
where coarsening is curvature-driven. In contrast, in the absence of
stickiness, the system belongs to the universality class of the
voter model, where coarsening is noise-driven.

In future work, it would be worthwhile investigating the
relationship between the ratio of the stickiness values, and the
critical value corresponding to the tipping point. Furthermore,
empirical data from venues like  massively multi-player online role
playing games \cite{Grabowski2009} could be used as a test bed for
validating our model and estimating the parameters which govern
inertial in opinion change. Lastly, controlled experiments with
incentives on online labor markets \cite{Suri2011} could further
narrow down the conditions under which stickiness becomes a
discernible feature of opinion dynamics.

\section*{Acknowledgments}
This work was supported in part by the Army Research Laboratory
(ARL) under Cooperative Agreement Number W911NF-09-2-0053, by the
Army Research Office (ARO) grant W911NF-12-1-0546, by the Office of
Naval Research (ONR) Grant No. N00014-09-1-0607, and by the National
Science Foundation (NSF) Grant No. DMR-1246958. The views and
conclusions contained in this document are those of the authors and
should not be interpreted as representing the official policies
either expressed or implied of the Army Research Laboratory or the
U.S. Government.

\pagebreak
%\bibliography{StickyOpinions}

\providecommand{\newblock}{}

\end{document}